\documentclass{elsart1p}

\usepackage{epsfig}
\usepackage{amsfonts}
\usepackage{bbm}

\newcommand{\ba}{\begin{eqnarray}}
\newcommand{\ea}{\end{eqnarray}}
\newcommand{\vc}[1]{{\mathbf #1}}
\newcommand{\redq}{{q}}

\newcommand{\tinymsbar}{{\overline{\mbox{\tiny\rm{MS}}}}}
\renewcommand{\(}{\left(}
\renewcommand{\)}{\right)}

\newcommand{\eq}{Eq.~}
\newcommand{\fig}{Fig.~}
\newcommand{\se}{Sec.~}
\newcommand{\nr}[1]{(\ref{#1})}

\begin{document}

\begin{frontmatter}



\hfill{BI-TP 2008/27} \\
\hfill{arXiv:0810.2718}

\vspace*{-8mm}

\title{Chasing electric flux in hot QCD}


\author{Y.~Schr\"oder}

\address{Faculty of Physics, University of Bielefeld, 
D-33501 Bielefeld, Germany}

\begin{abstract}
In this talk, I present the status of attempts to analyze the behavior of the 
so-called spatial 't~Hooft loop, which can be taken as an order parameter 
for the deconfinement phase transition in pure SU($N$) gauge theory. While 
lattice data show a strikingly universal scaling of extracted 
$k$\/--string tensions for various values of $k$ and $N$, 
the analytic approach to these observables might need some refinement.
\end{abstract}

\end{frontmatter}



\section{Introduction}

Four--dimensional SU($N$) gauge theory undergoes a phase 
transition from 
a cold and confined phase into a hot and deconfined phase
at a critical temperature $T_c$,
which is intimately related to the breaking of the Z($N$) centergroup
symmetry.
In this respect, a widely used order parameter is the Polyakov loop
$\,P = \frac1N\,{\rm Tr}{\cal P}
\exp(ig\int_0^{1/T}dx_0\,A_0)$,
whose thermal average $\langle P \rangle \neq 0$ in the deconfined phase.
There are however some formal issues associated with the use of 
$\langle P \rangle$ as order parameter:
$P$ cannot be defined at strictly zero temperature,
and its lattice-discretized version exhibits ultraviolet divergences 
in the continuum limit.

Analyzing other choices, and noting that the spatial Wilson loop shows 
only area law behavior at $T\neq0$, while at $T=0$ it exhibits 
area/perimeter law behavior for broken/unbroken Z($N$) symmetry,
it has been proposed to use the 't~Hooft flux loop operator 
\cite{'tHooft:1977hy} -- which in some sense 
is dual to Wilson loops -- as an alternative 
order parameter \cite{Kovner:1992fm}.

In order to give a definition of the 't~Hooft flux loop operator,
consider measuring color-electric flux through a large surface $S$,
in SU($N$) gauge theory at finite temperature.
Since the color-electric field $(\vc E)_i=F_{0i}$ is in the 
Lie Algebra of SU($N$), one asks for its flux projected onto some 
special direction $Y$ in SU($N$):
$\Phi=\frac1g \int_S d\vc S\cdot {\rm Tr}\vc E Y$.
Note that $\langle \Phi \rangle=0$.
Now define the 't~Hooft loop (or E-flux) 
operator \cite{'tHooft:1977hy,Kovner:1992fm}
$V=\exp(4\pi i\, \Phi)$,
whose thermal average $\langle V \rangle$ behaves non-trivially.

The directions (in matrix space) $Y$ are constrained by gauge invariance.
Since $F_{\mu\nu}^\Omega=\Omega F_{\mu\nu} \Omega^{-1}$ transforms as an 
adjoint, one has to choose special $Y$'s, namely generators of the center 
of SU($N$) which respect
$\exp(2\pi i Y_k)=\exp(2\pi i \frac{k}N)\,1\!\!\!\!\!\,1_{N\times N}$ 
for $k=0,...,N-1$.
These are not unique; one simple choice \cite{KorthalsAltes:2000gs} is
$Y_k=\frac1N\,{\rm diag}(\{k\}^{N-k},\{k-N\}^k)$.
One hence obtains a set of gauge invariant 
(for a detailed proof\footnote{To prove that $V_k$ acting 
on physical states is gauge invariant, the main idea is
to construct the physical subspace spanned by gauge invariant
combinations of gauge fields
(which are Wilson loops 
$W={\rm Tr}{\cal P}\exp(i\oint_C d\vc l\cdot A)$), 
to show  
$V_k\,W\,V_k^\dagger=\exp(2\pi i Y_k)\,W$
and therefore
$V_k^\Omega\,W\,(V_k^\Omega)^\dagger=\exp(2\pi i \Omega Y_k \Omega^{-1})\,W=
\Omega\,\exp(2\pi i Y_k )\,\Omega^{-1}\,W=\exp(2\pi i Y_k)\,W$,
such that when acting on a physical state constructed from 
one Wilson loop one obtains
$V_k^\Omega |{\rm phys}\rangle=
V_k^\Omega\,W\,(V_k^\Omega)^\dagger\,V_k^\Omega|0\rangle$.} 
see \cite{CKAwriteup}) 
operators 
\ba
V_k=\exp\(\frac{4\pi i}g \int_S d\vc S\cdot {\rm Tr}\vc E Y_k\)\;.
\ea
One can finally define the thermal averages 
\ba
\label{eq:sig}
\langle V_k \rangle \sim \exp\(-\sigma_k(T) A_S\)\;,
\ea
which at large areas $A_S$ of the surface $S$ 
define the $k$\/--string tensions $\sigma_k$.

\begin{figure}[t]
\centerline{\epsfysize=6.0cm\epsfbox{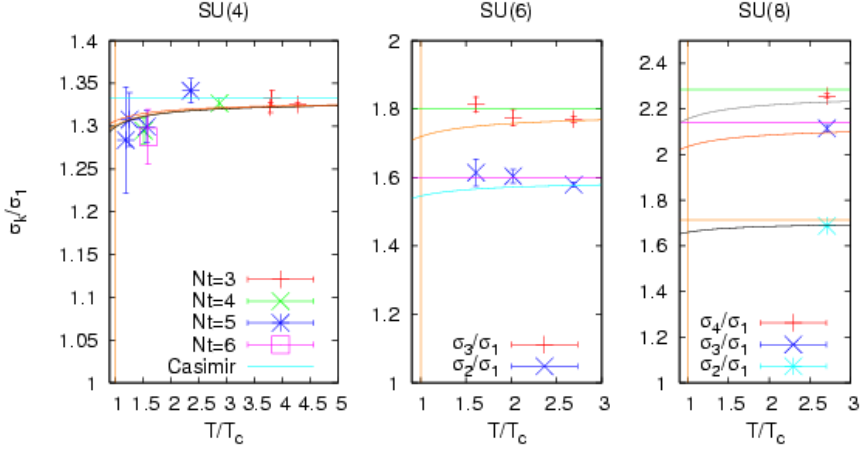}}
\caption[a]{Comparison of lattice data and analytical results 
for ratios of the $k$\/--string tensions of 
\eq\nr{eq:sig} as a function of temperature, 
for different gauge groups. 
The horizontal lines are analytic leading-order (LO) results 
$\frac{k(N-k)}{(N-1)}$, the curved lines are NNLO with 
$T_c/\Lambda_{\tinymsbar}=(1.1,1.35)$.
The plots are from ref.~\cite{deForcrand:2005rg}.}
\label{fig:sun22}
\end{figure}

The tensions $\sigma_k$ can be computed
either with analytic methods in a weak-coupling expansion 
(see \se\nr{sec:analytic}),
or on the lattice \cite{deForcrand:2005rg}.
From \fig\nr{fig:sun22},
one observes fairly good agreement between analytic and numerical results
for the dimensionless ratios $\sigma_k/\sigma_1$.

Interestingly, when choosing a different normalization on the 
vertical axis, on the lattice
one clearly observes more violation of ``Casimir scaling'' at low 
temperature \cite{deForcrand:2005rg}, as shown in \fig\nr{fig:su_all}. 
Moreover, and most strikingly, the lattice data for different gauge 
groups and different values of $k$ appear to fall onto a universal 
curve, hinting towards (large--$N$?) universality.
Clearly, it would be nice to understand this behavior analytically.

\begin{figure}[t]
\centerline{\epsfysize=6.0cm\epsfbox{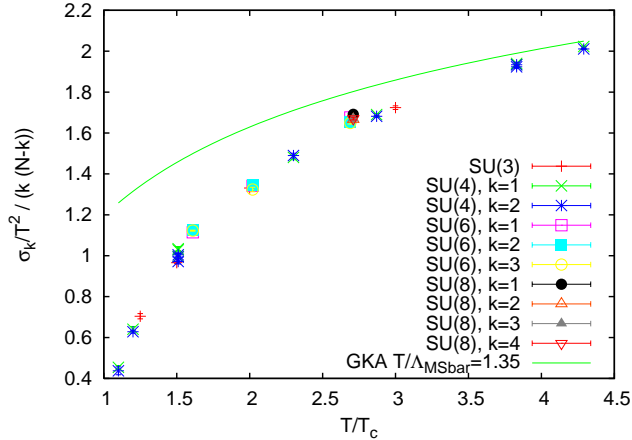}}
\caption[a]{The same data as in \fig\nr{fig:sun22}, with a different 
normalization. Note the universality for different $N$, $k$.
Solid line: NLO $\sigma_k$ with 2-loop running coupling, 
at $T_c/\Lambda_{\tinymsbar}=1.35$. 
The plot is taken from ref.~\cite{deForcrand:2005rg}.}
\label{fig:su_all}
\end{figure}

\section{Analytic results}
\label{sec:analytic}

The weak-coupling expansion has been driven to next-to-next-to 
leading order (NNLO),
\ba 
\hspace*{-5mm}\frac{\sigma_k}{T^2}=\frac{4\pi^2}{3\sqrt{3 g^2 N}}\,k(N-k)
\(1-15.2785..\frac{g^2 N}{(4\pi)^2}+ c_3(N,k) \frac{g^3N^{\frac32}}{(4\pi)^3}
+c_4(N,k) \frac{g^4 N^2}{(4\pi)^4}\) .
\ea
At leading order, one observes Casimir scaling,
while the NLO coefficient \cite{Bhattacharya:1990hk} scales in the same
way with $N$ and $k$, provided the expansion is organized in terms of $g^2 N$.

The NNLO coefficient $c_3(N,k)$ has been determined in 
\cite{Giovannangeli:2002uv}. 
It is small
(which is the reason why, for simplicity,  
in \fig\nr{fig:su_all} only NLO is shown); 
it contributes (in analogy to e.g. weak-coupling 
results for the QCD pressure \cite{Kapusta:1979fh}) 
with a sign opposite to the $g^2$ term; 
its nonanalytic (in $\alpha_{\rm s}$) magnitude $g^3$
is well-understood, originating from infrared sensitive 
${\cal O}(g^4)$ graphs that necessitate an all-order resummation 
of this specific sector;
and it is expressed in \cite{Giovannangeli:2002uv} as a specific 
two--dimensional integral plus an infinite sum.

From the analytic side, the question now clearly is whether the
unknown NNNLO term $c_4(N,k)$ can be computed, and whether it will 
help in explaining the apparent discrepancy at low temperatures 
$T\le 4T_c$ between lattice and analytic results as shown in
\fig\nr{fig:su_all}. To set our minds for doing this, it might be useful
to recall the setup in which the analytic results that are known so far 
have been obtained.

The basic idea is to compute $\sigma_k$ as tunneling effect 
through a perturbatively calculable potential barrier. 
Hence, one first needs to compute an effective potential.
The presence of $V_k$ breaks the center group symmetry, since
it chooses specific directions $Y_k$.
This leads to the Polyakov loop $P$ picking up a phase 
$\exp(2\pi i \frac{k}N)$ when moving through the surface $S$ 
(at, say, $z=0$, if the $S$ is simply chosen as spanned by an $L_x\times L_y$
loop in the $xy$ plane), which means that the field component 
$A_0(z)$ is discontinuous at $S$.
One can parameterize this behavior with (diagonal, traceless, $N\times N$)
matrices $C$, according to 
$P\equiv{\rm Tr}\,\exp(iC(z))$.
Now, constraining the path integral to these allowed values of $A_0$
with a delta function one obtains
\ba
\langle V_k \rangle &=& \int[{\cal D}A_0]
[{\cal D}\vc A] \,\delta\(P-{\rm Tr}\,\exp(iC(z))\) \,\exp(-S[A])
\;\sim\; \exp(-L_xL_y\,\sigma_k)
\;,
\nonumber
\ea
where the last step is valid in the limit of 
large loops and defines the minimum value of the effective
potential $U$, with $\sigma_k={\rm min}_{\{C(z)\}}\,U[C(z)]$.
This minimization is taken with respect to all $C(z)$ which have 
a $2\pi k$ discontinuity at $z=0$ and vanish as $z\rightarrow\infty$.

The minimal profile of the effective potential $U$ is conjectured to
be realized along the simplest path
in the space of the matrices $C\sim 2\pi \redq Y_k$, with $0<\redq<1$
(while there is no general proof yet, this has been shown at large $N$ and 
for $N=3,4$).
To compute $U$ in a perturbative expansion around the Polyakov 
loop constraint, one sets $A_0=C(z)+gQ_0$ and $A_i=g Q_i$,
chooses background field gauge
$S_{\rm gf}=\frac1\xi {\rm Tr}[D_\mu(C) Q_\mu]^2$,
performs the loop expansion, 
and finally minimizes with respect to $\redq$.
On the technical side, the $\redq$\/-dependence represents a smooth 
interpolation between bosonic and fermionic Matsubara 
sums \cite{Schroder:2008ex}. Additionally, 
going beyond NLO entails resumming the IR enhanced $g^4$ diagrams, 
which is most transparently done in the framework of dimensional reduction.
As the magnetic sector does not contribute before $g^5$,
the relative order $g^4$ is computable.

\section{Conclusions}

The spatial 't~Hooft loop $V_k$ measures the color-electric flux,
which at large temperatures is due to free screened gluons.
This induces an area law behavior, whose corresponding tensions 
$\sigma_k \sim \ln \langle V_k \rangle$ 
are hard to measure on the lattice, since they are exponentially suppressed,
and hence trigger a sampling problem that increases with the loop size.
Existing lattice measurements indicate universal behavior of the $\sigma_k$,
while NNLO weak-coupling results seem to correctly predict ratios 
of $\sigma_k$, but not their normalization close to $T_c$.
Hope is expressed that knowledge of the NNNLO term
would cure this lack of predictivity, and the setup in which to
obtain this contribution is roughly sketched.
As a final remark, let us stress that the ratios $\sigma_k/\sigma_1$ 
can serve as stringent tests for formulations of Yang--Mills theory 
derived from e.g. string theory.


\end{document}